# Stochastic Cooling with Strong Band Overlap

Valeri Lebedev[1]
Fermilab, PO box 500, Batavia, IL 60510, USA

**Abstract:** Up to present time the stochastic cooling was only tested and used at the microwave frequencies. Majority of these stochastic cooling systems operate without Schottky band overlap which greatly simplifies tuning of cooling systems and removes unwanted coupling between different cooling systems. A transition from the microwave stochastic cooling to the optical stochastic cooling or to the coherent electron cooling increases the central frequency of cooling systems by orders of magnitude and makes impossible a cooling system operation without overlap of Schottky bands. In this paper we consider how the band overlap affects the maximum cooling rate and the optimal gain.

## Introduction

Up to present time the stochastic cooling systems were only used at the microwave frequencies, typically in the range 1 to 8 GHz. In this case the slip-factor of a storage ring can be adjusted so that to avoid an overlap of Schottky bands. It greatly simplifies tuning of cooling systems and removes unwanted coupling between different cooling systems. As will be seen below an operation with the optimal slip-factor, where the Schottky bands just started to be overlapped, results in the fastest cooling. Further increase of band overlap reduces the cooling rates. It can also create additional problems with system tuning and operation.

With a transition from the microwave stochastic cooling [1,2] to the Optical Stochastic Cooling (OSC) [3] or to the Coherent Electron Cooling (CEC) [4], which operate at orders of magnitude higher frequencies, a cooling system operation without band overlap becomes impossible. In this paper we consider how band overlap effects on the optimal gain and the cooling rate. First, we consider cooling of continuous beam, and, then, analyze cooling of a bunched beam.

Our consideration is based on the general stochastic cooling theory [5,6] which is equally applicable for cooling with and without band overlap. In the course of discussion we will consider corrections specific for the OSC and CEC which appear due to much higher frequency. In this paper we will look into longitudinal cooling only. Corresponding equations for transverse cooling can be obtained by following the developed below procedure applied to the equations for transverse cooling derived in Ref. [6].

There are three major methods of longitudinal stochastic cooling [1,2]. They are the Palmer cooling, the filter cooling and the transit-time cooling. Only the latter can be used at optical frequencies. Here we consider the Palmer cooling and the transit-time cooling, and do not consider the filter cooling because it cannot operate with band overlap.

For a continuous beam an evolution of longitudinal distribution function, $\psi(x,t)$, in the course of stochastic cooling is determined by the Fokker-Planck equation [5]:

$$\frac{\partial \psi(x,t)}{\partial t} + \frac{\partial}{\partial x}\left(F(x)\psi(x,t)\right) = \frac{1}{2}\frac{\partial}{\partial x}\left(D(x)\frac{\partial \psi(x,t)}{\partial x}\right) \ , \quad (1)$$

where $x \equiv \Delta p/p$ is the relative momentum deviation, $t$ is the time,

$$F(x) = \frac{1}{T_0}\sum_{n=-\infty}^{\infty}\frac{G(x,\omega_n(x))}{\varepsilon(\omega_n(x))}e^{2\pi i n\eta_{pk}x} \quad (2)$$

is the cooling force,

$$D(x) = \frac{N}{T_0}\sum_{n=-\infty}^{\infty}\frac{|G(x,\omega_n(x))|^2}{|\varepsilon(\omega_n(x))|^2}\sum_{k=-\infty}^{\infty}\frac{1}{|k\eta|}\psi\left(\frac{k-(1-\eta x)n}{k\eta}\right) \quad (3)$$

is the diffusion,

---

* Work supported by Fermi Research Alliance, LLC under Contract No. De-AC02-07CH11359 with the United States Department of Energy.
1) Email: val@fnal.gov

$$\varepsilon(\omega) = 1 + N \int_{\delta \to 0_+} \frac{d\psi(x)}{dx} \frac{G(x,\omega)e^{i\omega T_0 \eta_{pk} x}}{e^{i\omega T_0(1+\eta x)} - (1-\delta)} dx \quad (4)$$

is the dielectric function, $T_0$ is the revolution time for the reference particle, $N$ is the number of particles, $G(x,\omega)$ is the generalized gain function, $\omega_n(x) = n\omega_0(1-\eta x)$, $\omega_0 = 2\pi/T_0$ is the revolution angular frequency, $\eta = \alpha - 1/\gamma^2$ is the ring slip-factor, $\alpha$ is the ring momentum compaction, $\gamma$ is the relativistic factor, $\eta_{pk}$ is the partial pickup-to-kicker slip-factor determined so that the change of pickup-to-kicker flight time relative to the reference particle is equal to: $\Delta T_{pk} = T_0 \eta_{pk} x$, and $\psi$ is normalized so that $\int \psi(x) dx = 1$. Here we neglected in Eq. (3) the contribution of amplifier noise to the diffusion, which is well justified for most practical systems. The sum over $k$ in Eq. (3) accounts for the band overlap. In the absence of the overlap only one addend in the sum is left and the sum is equal to $\psi(x)/|n\eta|$. In the case of strong band overlap the Schottky noise does not depend on the frequency. Consequently, in the cases of no band overlap and the strong band overlap Eq. (3) can be reduced to:

$$D(x) = \frac{N}{T_0} \begin{cases} \sum_{n=-\infty}^{\infty} \frac{|G(x,\omega_n(x))|^2}{|\varepsilon(\omega_n(x))|^2} \frac{\psi(x)}{|n\eta|}, & \text{no band overlap,} \\ \sum_{n=-\infty}^{\infty} |G(x,\omega_n(x))|^2, & \text{strong band overlap.} \end{cases} \quad (5)$$

Here in the bottom equation we accounted that $\varepsilon(\omega) \approx 1$ which is well justified for a system operating with strong band overlap at the optimal gain.

Multiplying Eq. (1) by $x^2$ and integrating over $x$ one obtains an equation describing evolution of rms momentum spread:

$$\frac{d\overline{x^2}}{dt} = -2\frac{G}{G_{ref}}\overline{F} + \left(\frac{G}{G_{ref}}\right)^2 \overline{D}, \quad (6)$$

where the integrals of cooling force and diffusion,

$$\overline{F} = -\int x F(x)\psi(x) dx,$$
$$\overline{D} = \int \psi(x) \frac{d}{dx}(xD(x)) dx, \quad (7)$$

are calculated for an arbitrary reference gain $G_{ref}$. As one can see the first term in the right side of Eq. (6) is proportional to $G$ and the second one counteracts cooling and is proportional to $G^2$. It implies that there is optimal value of the gain where the cooling rate achieves its maximum. Equaling a derivative of Eq. (6) over $G$ to zero one obtains the optimal gain:

$$G_{opt} = G_{ref} \frac{\overline{F}}{\overline{D}}, \quad (8)$$

and the maximum rate of momentum cooling:

$$\left.\frac{d\overline{x^2}}{dt}\right|_{max} = \frac{\overline{F}^2}{\overline{D}}. \quad (9)$$

Note that the maximum rate of momentum cooling does not depend on the choice of $G_{ref}$.

## 1. Palmer Cooling with Band Overlap

First, we consider the Palmer cooling of continuous beam. The gain function for the Palmer cooling can be presented in the following form: $G(x,\omega) = -xG'(\omega)$. In further calculations we assume the Gaussian distribution over momentum[1],

$$\psi(x) = \frac{1}{\sqrt{2\pi}\sigma_p} \exp\left(-\frac{x^2}{2\sigma_p^2}\right), \quad (10)$$

and a perfectly phased rectangular band: $G'(\omega) = G'$ for $\omega \in [\omega_1, \omega_2]$ and zero otherwise; $\text{Im}(G') = 0$. We also set $\varepsilon(\omega) = 1$. Accuracy of such approximation will be discussed later.

Using Eq. (2) one obtains the cooling force equal to:

$$F(x) = -\frac{G'}{\pi}(\omega_2 - \omega_1) x \Im(x),$$
$$\Im(x) = \frac{\sin(2\pi n_2 \eta_{pk} x) - \sin(2\pi n_1 \eta_{pk} x)}{2\pi(n_2 - n_1)\eta_{pk} x}, \quad (11)$$

where $n_1 = \omega_1/\omega_0$ and $n_2 = \omega_2/\omega_0$ are the harmonic numbers at the boundaries of cooling band. The form-factor $\Im(x)$ reduces the cooling force for large momentum deviations. The momentum deviation where the cooling force

---

[1] Strictly speaking the distribution function does not stays Gaussian in the course of beam cooling. Although a usage of Gaussian distribution yields compact and clear results an accuracy of such approximation depends on the details of cooling process and needs to be separately analyzed.

approaches zero the first time,

$$x_b = \frac{1}{2(n_2+n_1)|\eta_{pk}|}, \quad (12)$$

determines the cooling range. Outside of the cooling range $[-x_b, x_b]$ the cooling is replaced by heating. The partial slip-factor $\eta_{pk}$ has to be sufficiently small so that cooling would be present in the required momentum aperture. If possible, its zero value is desirable. Fig. 1 shows behavior of cooling form-factor on the momentum for different ratios of upper and lower band boundaries.

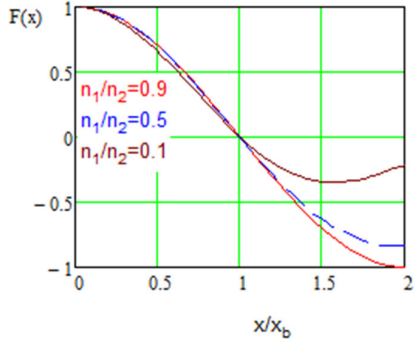

Fig. 1: Dependence of cooling force form-factor $\Im(x)$ on particle momentum for Palmer cooling.

In the below estimate of cooling rate we assume that all core particles are located far away from the cooling boundaries. In this case we can assume that $\Im(x)=1$ and, consequently, the cooling force is linear. Using Eq. (7) one finds the cooling force integral:

$$\bar{F} = 2G'(n_2 - n_1)\frac{\sigma_p^2}{T_0}. \quad (13)$$

Using Eq. (3) we obtain the diffusion:

$$D(x) = \frac{2NG'^2 x^2}{T_0} \sum_{n=n_1}^{n_2} \sum_{k=-\infty}^{\infty} \frac{1}{|k\eta|} \psi\left(\frac{k-(1-\eta x)n}{k\eta}\right). \quad (14)$$

Substituting this equation to Eq. (7), using the distribution function of Eq. (10) and performing integration over momentum one obtains the diffusion integral:

$$\bar{D} = \frac{3NG'^2 \sigma_p}{4\sqrt{\pi} T_0} \sum_{n=n_1}^{n_2} \sum_{k=-\infty}^{\infty} \left[1+4n^2 S_{kn}^2 + \frac{4}{3}n^4 S_{kn}^4\right] \\ \times \frac{e^{-(2k^2-n^2)S_{kn}^2}}{|k\eta|}, \quad S_{kn} = \frac{(k-n)}{2k^2\eta\sigma_p}. \quad (15)$$

Taking into account that the bandwidth of single Schottky band is much smaller than the bandwidth of the cooling system we can neglect difference between $k$ and $n$ everywhere except their direct difference. That yields:

$$\bar{D} = \frac{3NG'^2 \sigma_p}{4\sqrt{\pi} T_0} \times \\ \sum_{n=n_1}^{n_2} \sum_{k=-\infty}^{\infty} \left[1+\left(\frac{k}{\eta\sigma_p n}\right)^2 + \frac{1}{12}\left(\frac{k}{\eta\sigma_p n}\right)^4\right] \frac{e^{-\left(\frac{k}{2\eta\sigma_p n}\right)^2}}{|k\eta|}. \quad (16)$$

To find the sum over $k$ we introduce a function

$$\Phi_{PC}(x) = \sum_{k=-\infty}^{\infty} \left(1 + 4\frac{k^2}{x^2} + \frac{4}{3}\frac{k^4}{x^4}\right) e^{-k^2/x^2}. \quad (17)$$

Then, the diffusion integral can be written in the following form:

$$\bar{D} = \frac{3NG'^2 \sigma_p}{4\sqrt{\pi} T_0} \int_{n_1}^{n_2} \frac{\Phi_{PC}(2\eta\sigma_p u)}{|u\eta|} du, \quad (18)$$

where we replaced summing over $n$ by an integration over $u$. It presents a good approximation due to very large values of $n_1$ and $n_2$.

A plot of $\Phi_{PC}(x)$ is presented in Fig. 2. The function has the following asymptotic behavior:

$$\Phi_{PC}(x) = \begin{cases} 1, & x \ll 1, \\ 4\sqrt{\pi} x, & x \gg 1, \end{cases} \quad (19)$$

and can be approximated within 3% by the following function:

$$\Phi_{PC}(x) \approx \frac{1}{1+x^6} + 4\sqrt{\pi} x e^{-1/(18x^4)}. \quad (20)$$

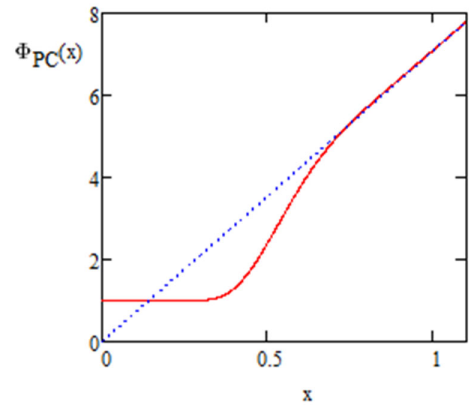

Fig. 2: Plot of function $\Phi_{PC}(x)$. Blue dashed line presents asymptotic for large $x$.

Substituting asymptotes presented in Eq. (19) to Eq. (18) one obtains the diffusion integral without and with strong band overlap (top and bottom, respectively):

$$\bar{D} \approx \frac{3NG'^2}{T_0} \begin{cases} \dfrac{\sigma_p}{4\sqrt{\pi}\,|\eta|} \ln\left(\dfrac{n_2}{n_1}\right), & 2\eta\sigma_p n_2 \ll 1, \\ 2\sigma_p^2 (n_2 - n_1), & 2\eta\sigma_p n_1 \gg 1. \end{cases} \quad (21)$$

Using Eq. (9) we obtain the cooling rate of the rms momentum spread for the Palmer cooling of continuous beam[2]:

$$\lambda_{max} = \frac{1}{\sigma_p^2} \left.\frac{d\overline{x^2}}{dt}\right|_{max} = \frac{\overline{F^2}}{\sigma_p^2 \overline{D}} \quad . \quad (22)$$

Figure 3 presents the dependence of dimensionless cooling rate on the value of band overlap for systems with different bandwidths ($n_2/n_1$=1.1, 1.5, 2). As one can see the maximum cooling rate, $\lambda \approx 4W/(3N)$, is achieved at a minor band overlap, $\eta\sigma_p n_2 \approx 0.2$. Here $W=(n_2-n_1)/T_0$ is the bandwidth of the cooling system. Further increase of band overlap reduces the cooling rate by about factor of 2. Using asymptotes of Eq. (21) and Eqs. (22) and (13) one obtains the maximum cooling rates in the absence of band overlap and with strong band overlap (top and bottom, respectively):

$$\lambda_{max} \approx \frac{2W}{3N} \begin{cases} \dfrac{8\sqrt{\pi}(n_2-n_1)|\eta|\sigma_p}{\ln(n_2/n_1)}, & 2\eta\sigma_p n_2 \ll 1, \\ 1, & 2\eta\sigma_p n_1 \gg 1. \end{cases} \quad (23)$$

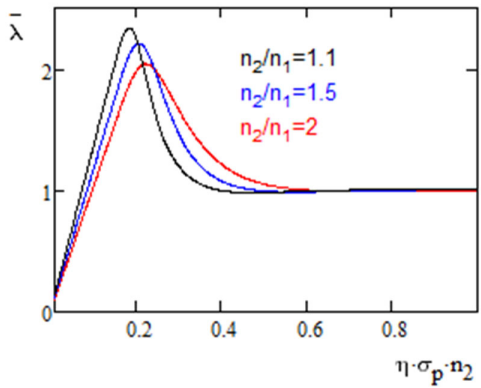

Fig. 3: Dimensionless cooling rate, $\bar{\lambda} = 3NT_0\lambda/(2W)$, for Palmer cooling at the optimal gain as a function of the band overlap at the upper band end for different bandwidths of the system.

---

[2] In this paper all cooling rates are calculated for the square of rms momentum spread or the longitudinal

In the above calculations we neglected that the dielectric function is different from 1. As it is pointed out in Ref. [6] it makes comparatively small correction. Such behavior is associated with the following. Both $\overline{F^2}$ and $\overline{D}$ are reduced by signal suppression in almost the same proportion: $\overline{F}^2 \propto \mathrm{Re}(1/\varepsilon)^2$ and $\overline{D}^2 \propto 1/|\varepsilon|^2$. For a well-phased system the difference in attenuations for both terms is comparatively small. The effect of signal suppression is more pronounced for a systems operating far from the band overlap. However, as it was pointed out in Ref. [6] (see page 304), even in this case it makes less than 5% correction to the cooling rate for one-octave system operating at the optimal gain, which implies the maximum value of $|\varepsilon|$ being about 2 and, consequently, the strong signal suppression for both the cooling force and the diffusion. Note that the effect of signal suppression can be much larger for a poorly phased cooling system. Finally we note that for a system operating with strong band overlap the signal suppression is small and can be completely neglected.

As one can see Eq. (23) describes the cooling which can cool the beam to the zero momentum spread. It is related to the approximation used in the model where we neglected all other diffusion mechanisms except the beam driven diffusion produced by cooling system. It resulted in that for small $x$ the diffusion is proportional to $x^2$ and is equal to zero for $x = 0$. In the real world there are other diffusion mechanisms (like intrabeam scattering, or thermal noise of cooling system) which limit the reduction of momentum spread. For the case of cooling with strong band overlap we can assume that the diffusion is: $D = D_0 + D_p x^2$, where $D_p$ is determined by Eq. (5) and $D_0$ is the external diffusion. Taking into account that the cooling force for small momentum deviations is linear, $F_p = -F_p x$, one obtains the Fokker-Plank equation:

$$\frac{\partial \psi}{\partial t} = \frac{\partial}{\partial x}\left(F_p x \psi\right) + \frac{1}{2}\frac{\partial}{\partial x}\left(\left(D_0 + D_p x^2\right)\frac{\partial \psi}{\partial x}\right) \quad . \quad (24)$$

Here both $F_p$ and $D_p$ are determined by the cooling system and its gain. At the optimal gain they are related as following: $D_p = F_p/3$; and the damping rate is equal to:

emittance. Note that the cooling rate for the momentum spread or longitudinal bunch size is twice smaller.

$\lambda_{max} = F_p$. It is straightforward to integrate Eq. (24) for the stationary state. The result is:

$$\psi_s(x) = \frac{8}{3\pi}\sqrt{\frac{\lambda_{max}}{3D_0}}\frac{1}{\left(1+\lambda_{max}x^2/(3D_0)\right)^3} \quad . \quad (25)$$

The corresponding rms momentum spread is equal to:

$$\sigma_s = \sqrt{D_0/\lambda_{max}} \; .$$

Now let us consider how the optimal gain and the maximum cooling rate are changed for a bunched beam cooling. The cooling force integral of Eq. (13) is not affected by the bunching but the diffusion integral is, and its value is increased. Replacing the longitudinal particle density of continuous beam, $N/T_0$, in Eq. (14) by the longitudinal density of bunched beam,

$$\left(N/(\sqrt{2\pi}\sigma_t)\right)\exp(-t^2/2\sigma_t^2) \; ,$$

and performing averaging with longitudinal distribution we obtain the diffusion integral for the bunched beam:

$$\bar{D}_b = \bar{D}\int_{-\infty}^{\infty}\frac{T_0 e^{-t^2/2\sigma_t^2}}{\sqrt{2\pi}\sigma_t}\frac{e^{-t^2/2\sigma_t^2}dt}{\sqrt{2\pi}\sigma_t} = \frac{T_0}{2\sqrt{\pi}\sigma_t}\bar{D} \quad . \quad (26)$$

Here $\sigma_t$ is the rms bunch duration, and $\bar{D}$ is the diffusion integral computed for a continuous beam with the same number of particles.

To obtain the cooling rate for a bunched beam one also has to take into account that in the linear RF the synchrotron motion reduces both the single particle cooling rate and the diffusion by two times. The single particle cooling rate is reduced because the average momentum deviation squared is reduced by two times due to synchrotron motion. The effect of diffusion is reduced because the beam heating introduced by diffusion is equally split between two degrees of freedom of oscillatory synchrotron motion (or, in other words, between potential and kinetic energies). We also need to note that for a Gaussian distribution the shape of momentum distribution does not depend on the longitudinal coordinate inside the bunch. Therefore it is decoupled from the longitudinal coordinate, and should not be additionally accounted in averaging along the bunch. Thus, to obtain the cooling rate of bunched beam one has to multiply the cooling rates of continuous beam presented in Eqs. (22) and (23) by factor $\sqrt{\pi}\sigma_t/T_0$.

## 2. Transit-time Cooling with Band Overlap

In transition from microwave frequencies to optical frequencies we lose an ability to create difference signals used in the Palmer cooling. In this case the transit-time cooling is the only practical choice. The gain of such system is: $G(x,\omega) = -iG(\omega)$. Then, for a continuous beam the cooling force of a system with rectangular band ($G'(\omega) = iG_0$ for $\omega \in [\omega_1, \omega_2]$ and zero otherwise) is:

$$F(x) = \frac{2G_0}{T_0}\sum_{n=n_1}^{n_2}\text{Im}\left(e^{2\pi i n \eta_{pk}x}\right) = \frac{2G_0}{T_0}\frac{\sin(\pi\eta_{pk}(n_2-n_1)x)}{\pi\eta_{pk}x}\sin(\pi\eta_{pk}(n_2+n_1)x). \quad (27)$$

Corresponding cooling range is:

$$x_b = \frac{1}{\eta_{pk}(n_2+n_1)} \quad . \quad (28)$$

For small amplitude particles the cooling force is obtained by expending Eq. (27). It is equal to:

$$F(x) = \frac{2\pi\eta_{pk}G_0}{T_0}\left(n_2^2 - n_1^2\right)x \quad . \quad (29)$$

Substituting it to Eq. (7) we obtain the force integral:

$$\bar{F} = \frac{2\pi\eta_{pk}G_0}{T_0}\left(n_2^2 - n_1^2\right)\sigma_p \quad . \quad (30)$$

Similar to the previous section we find the diffusion,

$$D(x) = \frac{2NG_0^2}{T_0}\sum_{n=n_1}^{n_2}\sum_{k=-\infty}^{\infty}\frac{1}{|k\eta|}\psi\left(\frac{k-(1-\eta x)n}{k\eta}\right) \; , \quad (31)$$

and the diffusion integral,

$$\bar{D} = \frac{NG_0^2}{2\sqrt{\pi}T_0\sigma_p}\sum_{n=n_1}^{n_2}\sum_{k=-\infty}^{\infty}\left[1+\frac{1}{2}\left(\frac{k}{\eta\sigma_p n}\right)^2\right]\frac{e^{-k^2/(2\eta\sigma_p n)^2}}{|k\eta|} \; , \quad (32)$$

where following the procedure described in the previous section we only left significant terms in the difference between $k$ and $n$. To find the sum over $k$ we introduce function:

$$\Phi_{TC}(x) = \sum_{k=-\infty}^{\infty}\left(1+2\frac{k^2}{x^2}\right)e^{-k^2/x^2} \quad . \quad (33)$$

Then, the diffusion integral can be written in the following form:

$$\bar{D} = \frac{NG_0^2}{2\sqrt{\pi}T_0\sigma_p}\int_{n_1}^{n_2}\frac{\Phi_{PC}(2\eta\sigma_p u)}{|u\eta|}du \; . \quad (34)$$

A plot of $\Phi_{TC}(x)$ is presented in Fig. 4. The function has the following asymptotic behavior:

$$\Phi_{TC}(x) = \begin{cases} 1, & x \ll 1, \\ 2\sqrt{\pi}x, & x \gg 1, \end{cases} \quad (35)$$

and can be approximated with about 3% accuracy by the following function:

$$\Phi_{TC}(x) \approx \frac{1}{1+4x^6} + 2\sqrt{\pi}xe^{-1/(10.4x^4)}. \quad (36)$$

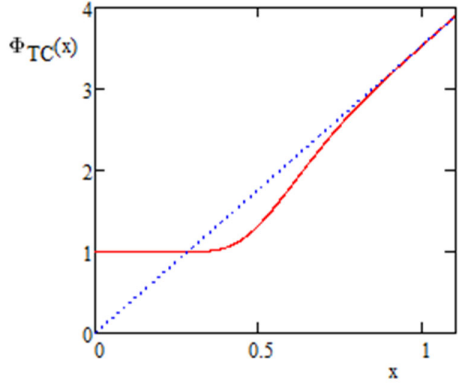

Fig. 4: Plot of function $\Phi_{TC}(x)$. Blue dashed line presents asymptotic for large $x$.

Substituting asymptotes presented in Eq. (35) to Eq. (34) one obtains the diffusion without and with strong band overlap:

$$\bar{D} \approx \frac{2NG_0^2}{T_0} \begin{cases} \dfrac{\sigma_p}{4\sqrt{\pi}|\eta|\sigma_p} \ln\left(\dfrac{n_2}{n_1}\right), & 2\eta\sigma_p n_2 \ll 1, \\ (n_2 - n_1), & 2\eta\sigma_p n_1 \gg 1. \end{cases} \quad (37)$$

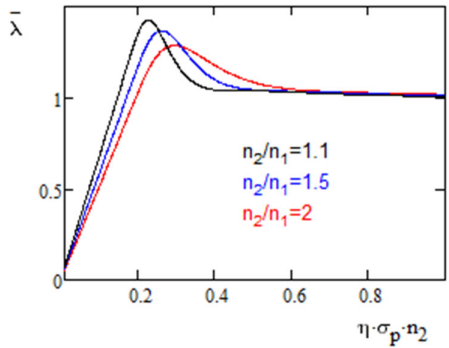

Fig. 5: Dimensionless cooling rate, $\bar{\lambda} = N x_{max}^2 \lambda / (2\pi^2 W \sigma_p^2)$, for transit-time cooling at optimal gain as a function of the band overlap at the upper band end for different bandwidths of the system.

Similarly, using Eq. (9) we obtain the cooling rate for the transit-time cooling. Figure 5 presents the dependence of dimensionless cooling rate on the value of band overlap for systems with different bandwidths ($n_2/n_1$=1.1, 1.5, 2). As one can see, the same as for the Palmer cooling, the maximum cooling rate, $\lambda_{max} \approx (W/N)(5.3\sigma_p/x_b)^2$, is achieved at a minor band overlap, $\eta\sigma_p n_2 \approx 0.25$. Using asymptotes of Eq. (35) we obtain the maximum cooling rate in the absence of band overlap and with strong band overlap (top and bottom, respectively):

$$\lambda_{max} \approx \frac{2\pi^2 \sigma_p^2}{x_b^2} \frac{W}{N} \begin{cases} \dfrac{4\sqrt{\pi}(n_2-n_1)|\eta|\sigma_p}{\ln(n_2/n_1)}, & 2\eta\sigma_p n_2 \ll 1, \\ 1, & 2\eta\sigma_p n_1 \gg 1. \end{cases} \quad (38)$$

In difference to the Palmer cooling the cooling rate of transit-time cooling is additionally reduced by the ratio $\sigma_p^2/x_b^2$. This difference is associated with much larger diffusion for small amplitude particles in the transit-time cooling due to different principles of signal processing in the Palmer cooling and the transit-time cooling. A particle contribution to the diffusion is suppressed proportionally to its momentum deviation squared ($\propto x^2$) in the Palmer cooling but is not suppressed in the transit-time cooling.

We note that in the case of transit-time cooling with strong band overlap an initial Gaussian distribution stays Gaussian for the entire time of cooling process. In this case, similar to Eq. (24) we obtain the following Fokker-Plank equation:

$$\frac{\partial \psi}{\partial t} = \frac{\partial}{\partial x}(F_p x \psi) + \frac{D_0}{2}\frac{\partial^2 \psi}{\partial x^2}. \quad (39)$$

Its solution is:

$$\psi(x,t) = \frac{1}{\sqrt{2\pi}\sigma(t)} \exp\left(-\frac{x^2}{2\sigma^2(t)}\right),$$
$$\sigma(t) = \sqrt{\frac{D_0}{2F_p} + \left(\sigma_0^2 - \frac{D_0}{2F_p}\right)\exp(-2F_p t)}. \quad (40)$$

Here $\sigma_0$ is the rms width of initial Gaussian distribution, and we took into account that the diffusion does not depend on $x$ for transit-time cooling with strong band overlap.

The same as for Palmer cooling the factor $\sqrt{\pi}\sigma_t/T_0$

accounts for a change in the cooling rate in the transition from continuous beam cooling to bunched beam cooling.

## 3. Optical Stochastic Cooling

The OSC [7] was suggested as a method to extend the frequency band of stochastic cooling from the microwave frequencies to the optical frequencies. It enables achieving stochastic cooling rates orders of magnitude faster than it could be achieved with microwave stochastic cooling. Due to very high frequency the OSC operates with very strong band overlap. In this section we consider how the described above theory of stochastic cooling can be applied to the OSC.

In the OSC a particle radiates electromagnetic wave in the first (pickup) wiggler. Then, this e.-m. wave is amplified and refocused to the second (kicker) wiggler as shown in Fig. 6. The particle beam is separated from the radiation by four-dipole chicane creating space for optical amplifier and optical lenses, and introducing a delay equal to a delay of radiation in the optical system, so that a particle would interact with its own radiation amplified by optical amplifier. We assume that the optical system compensates for the depth of field [8] and that the bandwidth of the amplifier is wider than the bandwidth of forward radiation. In this case a particle, which is shifted longitudinally by distance $s$ from the reference particle at the beginning of kicker wiggler, experiences the longitudinal kick in the kicker wiggler. Summed such kicks produce the cooling force equal to:

$$F(x) \equiv \frac{dx}{dt} \equiv \frac{d}{dt}\left(\frac{\Delta p}{p_0}\right) = -\frac{eE_0 K_u}{T_0 p_0 c \gamma} \times$$
$$\begin{cases} \int_{-(L_w/2-s)}^{L_w/2} \sin(k_w s')\cos(k_w s' - ks)ds', & 0 < \frac{ks}{k_w} < L_w, \\ \int_{-L_w/2}^{L_w/2-s} \sin(k_w s')\cos(k_w s' - ks)ds', & -L_w < \frac{ks}{k_w} < 0. \end{cases}$$
(41)

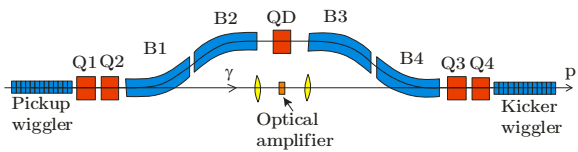

Fig. 6: Layout of the OSC system.

Here $E_0$ is the e.-m. wave amplitude, $\gamma$ is the particle relativistic factor, $p_0$ is the particle momentum, $L_w$ = $2\pi n_w / k_w$ is the wiggler length, $n_w$ is the number of wiggler periods, $k_w$ and $k$ are the wave numbers corresponding to the wiggler period and to the forward e.-m. radiation, respectively, $K_u$ is the undulator parameter, and $e$ is the particle charge. An integration in Eq. (41) yields:

$$F(x) = -F_0\left(1 - \left|\frac{ks}{2\pi n_w}\right|\right)\sin(ks), \quad -1 < \frac{ks}{2\pi n_w} < 1,$$
$$F_0 = \frac{eE_0 K_u L_w}{2T_0 p_0 c \gamma}$$
(42)

Here we additionally need to take into account that the particle longitudinal position in the kicker wiggler is related to its relative momentum deviation, $x$, by the following relationship:

$$s = C\eta_{pk} x \ .$$
(43)

where $C$ is the ring circumference. We need to note that the radiation frequency is decreased with an increase of angle between the radiation and the forward direction. One could expect that it results in a larger effective bandwidth of the cooling system. However, refocusing of radiation to the kicker undulator yields that the resulting e.-m. wave is summed from the waves which have the same period along the system axis independently from a wave direction. Therefore the limitation of the optical amplifier bandwidth affecting this off-axis radiation (having smaller frequency) weakly affects the shape of the e.-m. wave observed in the kicker wiggler, although it reduces the e.-m. wave amplitude. Fig. 7 presents a plot of cooling force for the 7-period wiggler.

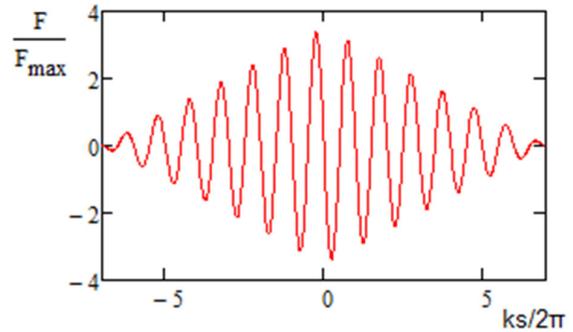

Fig. 7: Dependence of cooling force on the longitudinal particle position relative to the reference particle.

In transit-time cooling the gain function and the cooling force are related by the following relationships:

$$F(x) = \frac{1}{T_0} \sum_{n=-\infty}^{\infty} G_n e^{2\pi i \eta_{pk} n x} = \frac{1}{T_0} \int_{-\infty}^{\infty} G_n e^{2\pi i \eta_{pk} n x} dn, \quad (44)$$

$$G_n = \eta_{pk} T_0 \int_{-\infty}^{\infty} F(x) e^{-2\pi i \eta_{pk} n x} dx,$$

where we assume a very wide cooling band enabling a replacement of summation by integration. It also implies a strong band overlap, and, consequently, with good accuracy we can set in Eq. (2) $\varepsilon(\omega)=1$. Substituting Eq. (42) into the bottom equation of (44) one obtains the gain function:

$$G_n = -\eta_{pk} F_0 T_0 \int_{-2n_w x_b}^{2n_w x_b} \left(1 - \left|\frac{ks}{2\pi n_w}\right|\right) \sin(ks) e^{-2\pi i \eta_{pk} x n} dx, \quad (45)$$

where $x_b = \pi/(kC\eta_{pk})$ is the boundary momentum at which the cooling force is changing sign from cooling to heating (cooling range), and $s = C\eta_{pk} x$. Accounting the relationship of Eq. (43) and integrating one obtains the system gain:

$$G_n = \frac{iF_0 T_0 n_w}{2n_c} \times \left( \left(\frac{\sin(\pi n_w (n-n_c)/n_c)}{\pi n_w (n-n_c)/n_c}\right)^2 - \left(\frac{\sin(\pi n_w (n+n_c)/n_c)}{\pi n_w (n+n_c)/n_c}\right)^2 \right). \quad (46)$$

Here $n_c = kC/2\pi$ is the harmonic number in the band center.

Substituting this gain function into the bottom Eq. (5) and performing calculations we obtain for the diffusion and the diffusion integral:

$$\bar{D} = D = \frac{N F_0^2 T_0 n_w}{3 n_c}, \quad n_w \gg 1. \quad (47)$$

Here we assume that the number of wiggler periods is sufficiently large so that we could neglect interference between two addends in Eq. (47). We also took into account that $\int_{-\infty}^{\infty} (\sin^4(x)/x^4) dx = 2\pi/3$. For small amplitude particles Eq. (42) yields the cooling force and its integral:

$$F = \pi F_0 \frac{x}{x_b}, \quad (48)$$

$$\bar{F} = \pi F_0 \frac{\sigma_p^2}{x_b}.$$

Finally we obtain the maximum cooling rate for a continuous beam at the optimal gain:

$$\lambda_{\max} = \frac{\bar{F}^2}{\sigma_p^2 \bar{D}} = \frac{3\pi^2 f_c \sigma_p^2}{N n_w x_b^2}, \quad n_w \gg 1. \quad (49)$$

where $f_c = n_c/T_0$ is the central frequency of cooling system. Comparing this result with Eq. (38) one can see that the effective bandwidth of the system is $W = 3 f_c/(2n_w)$.

To obtain the maximum cooling rate for a bunched beam we follow the described above procedure. It results in:

$$\lambda_{\max} = \frac{3\pi^2 \sqrt{\pi} f_c}{N n_w} \frac{\sigma_t}{T_0} \frac{\sigma_p^2}{x_b^2}, \quad n_w \gg 1. \quad (50)$$

In the above consideration we assumed that the gain of cooling system does not change along the bunch. Typically, this condition cannot be fulfilled if a parametric optical amplifier [9] or a FEL is used as an optical amplifier. To simplify calculations we assume below that the gain dependence on the longitudinal coordinate related to the bunch does not depend on a harmonic number and can be approximated by a Gaussian function:

$$G_n(s) = G_n \exp\left(-\frac{s^2}{2\sigma_g^2}\right), \quad (51)$$

where $\sigma_g$ is the gain rms length. Then, the cooling force depends on the longitudinal coordinate as:

$$F(x,s) = \lambda_0 \exp(-s^2/(2\sigma_g^2)) x, \quad (52)$$

$$\lambda_0 = -\frac{4\pi \eta_{pk}}{T_0} \sum_{n=0}^{\infty} \text{Im}(G_n) n.$$

Here we expended the exponent in Eq. (2) for small amplitude oscillations. We also do not assume that the gain dependence on the harmonic number is described by Eq. (46). The cooling rate for a single particle with amplitude of synchrotron motion equal to $A$ is obtained by averaging over synchrotron period:

$$\frac{dA}{dt} = \lambda_0 \frac{A}{2\pi} \int_0^{2\pi} \cos^2(\phi) \exp\left(-\frac{A^2 \sin^2(\phi)}{2\sigma_g^2}\right) d\phi. \quad (53)$$

An integration results in:

$$\frac{dA}{dt} = -\frac{\lambda_0}{2} A \exp\left(-\frac{A^2}{4\sigma_g^2}\right) \left(I_0\left(\frac{A^2}{4\sigma_g^2}\right) + I_1\left(\frac{A^2}{4\sigma_g^2}\right)\right), \quad (54)$$

where $I_0(x)$ and $I_1(x)$ are the modified Bessel functions. Similarly, averaging over the entire bunch distribution one can write the cooling rate for the rms bunch length:

$$\frac{d}{dt}\overline{s^2} = -\frac{2\lambda_0}{\sigma_s^2}\int_0^{2\pi}\frac{d\phi}{2\pi}\times$$
$$\int_0^{2\pi} A^3 \cos^2(\phi)\exp\left(-\frac{A^2\sin^2(\phi)}{2\sigma_g^2}-\frac{A^2}{2\sigma_s^2}\right)dA,\quad (55)$$

where $A\sin(\phi)\equiv s$ accounts the dependence of gain on the longitudinal coordinate. To compute the integral, first, we integrate over $A$. Then, taking into account that,

$$\int_0^{2\pi}\frac{d\phi}{2\pi}\frac{\cos^2(\phi)}{\left(a+b\sin^2(\phi)\right)^2}=\frac{1}{2a^2\sqrt{1+b/a}},\quad (56)$$

we obtain:

$$\overline{F}_b \equiv \frac{d}{dt}\overline{s^2} = -\frac{\lambda_0 \sigma_s^2}{\sqrt{1+\sigma_s^2/\sigma_g^2}}.\quad (57)$$

Similarly, averaging the diffusion integral over bunch length can be written in the following form:

$$\overline{D}=\frac{D_0}{2}\int_{-\infty}^{\infty}\exp\left(-\frac{s^2}{2\sigma_g^2}\right)\left(\exp\left(-\frac{s^2}{2\sigma_s^2}\right)\right)^2\frac{ds}{\sqrt{2\pi}\sigma_s}.\quad (58)$$

Here in the second term in the integral (equal to the square of exponent) accounts for the distribution dependence on $s$ and the dependence of diffusion on $s$ due to reduced density of the particles. The factor ½ in front of integral accounts that the energy is distributed between two degrees of freedom, and

$$D_0 = \sqrt{\frac{2}{\pi}}\frac{N}{\sigma_t}\sum_{n=0}^{\infty}|G_n|^2,\quad (59)$$

is the diffusion computed in the bunch center with bottom equation of Eq. (5), where we replaced $N/T_0$ by $N/\sqrt{2\pi}\sigma_t$. Integrating Eq. (58) one obtains:

$$\overline{D}_b = \frac{D_0}{2\sqrt{2}}\frac{1}{\sqrt{1+\sigma_s^2/\left(2\sigma_g^2\right)}}.\quad (60)$$

Then, similar to the Eq. (22) we obtain the maximum cooling rate:

$$\lambda_b = \frac{\overline{F}_b^2}{4\sigma_p^2 \overline{D}_b}=F_g(\sigma_g,\sigma_s)\times$$
$$\frac{8\pi^{5/2}\eta_{pk}^2\sigma_p^2}{NT_0}\left(\sum_{n=0}^{\infty}n\,\mathrm{Im}(G_n)\right)^2\Big/\sum_{n=0}^{\infty}|G_n|^2,\quad (61)$$

where the fudge factor related to a finite duration of the gain is:

$$F_g(\sigma_g,\sigma_s)=\frac{\sigma_g^2}{\sigma_g^2+\sigma_s^2}\sqrt{1+\frac{\sigma_s^2}{2\sigma_g^2}}$$
$$=\begin{cases}1, & \sigma_g\gg\sigma_s,\\ \sigma_g/\left(\sqrt{2}\sigma_s\right), & \sigma_g\ll\sigma_s.\end{cases}\quad (62)$$

One can see that for the case $\sigma_g\ll\sigma_s$ the average cooling rate is reduced proportionally to the ratio of the gain length to the bunch length. Note, that in this case the cooling is concentrated in the bunch center and creates a bulb in the distribution center resulting in the distribution being far away from the Gaussian distribution which, subsequently, violates the approximation of Gaussian beam used for this estimate. Sweeping the location of the gain along the bunch can resolve this problem but it will result in additional reduction of the average cooling rate.

Note also that Eq. (61) is directly related to Eq. (38). In this comparison, one has to use the case $2\eta\sigma_p n_1 \gg 1$ in Eq. (38) and additionally account for the beam bunching; one also has to use an approximation of the rectangular band in Eq. (61). Also, Eq. (61) can be transformed to Eq. (50) if the OSC cooling force of Eq. (42) is used in Eq. (61).

## 4 Summary

In this paper, we considered how the Schottky band overlap affects the maximum cooling rates achievable in stochastic cooling. It was shown that the maximum cooling rate is achieved at small band overlap when $\eta\sigma_p n_2 \approx 0.2 \div 0.25$. With further increase of band overlap the cooling rate is decreased to a constant level and is not changed after $\eta\sigma_p n_2$ is above about 0.4, as it is shown in Figs. 3 and 5. Here $n_2$ is the harmonic number at the upper end of cooling band. In our consideration we used a Gaussian distribution for all calculations and neglected that the beam distribution does not necessarily stay Gaussian in the process of cooling. We also neglected all other diffusion mechanisms, like intrabeam scattering which accounting can be important for accurate description of beam distribution evolution.

It is important to note that in the consideration of Palmer cooling we neglected a contribution of betatron motion to the longitudinal diffusion. This effect is not present if the longitudinal and transverse Schottky bands do not overlap. Overlapping of longitudinal and transverse Schottky bands depends on the betatron tune and happens significantly earlier than the considered above overlap of longitudinal bands. In the case of Palmer cooling with strong band overlap the diffusion is increased in $(1+\sigma_\perp^2/\sigma_\parallel^2)$ times, where $\sigma_\perp$ and $\sigma_\parallel$ are the betatron and synchrotron rms beam sizes in the stochastic cooling pickup.

In the case of transit-time cooling, and, consequently, OSC the overall diffusion is much larger and additional diffusion due to transverse motion is insignificant.

With minor modifications the considered above estimates can be applied to the coherent electron cooling [4]. In this case one has to add to the diffusion a contribution coming from the longitudinal density perturbations in the electron bunch. Assuming that locations of electrons in the pickup (entrance of FEL where the cooled particles introduce perturbation to the electron beam) are random one obtains that the diffusion grows in $\approx \left(1+\left(N_e \sigma_s\right)/\left(N \sigma_g\right)\right)$ times and the cooling rate decreases in the same proportion. Here $N_e$ is the number of electrons in the electron bunch and we assume that the electron bunch length is equal to the gain length, $\sigma_g$.